\documentclass[
    a4paper,
    prb,
    bibnotes,
    twocolumn,
    showpacs,
    superscriptaddress,
    groupedaddress
]{revtex4}

\usepackage{array}
\usepackage{graphicx}
\usepackage{amssymb}
\usepackage{amsmath}
\usepackage{fullpage}
\usepackage{hyperref}
\usepackage{paralist}
\usepackage{subfigure}
\usepackage{float}
\hypersetup{
    colorlinks   = true,
     citecolor    = blue,
     linkcolor    = blue,
     citecolor    = blue,
     filecolor     = blue,
     urlcolor	      =	blue  
}

\setcitestyle{square}

\newcommand{\beq}{\begin{equation}}
\newcommand{\eeq}{\end{equation}}

\newcommand{\beqa}{\begin{eqnarray}}
\newcommand{\eeqa}{\end{eqnarray}}

\newcommand{\Eqref}[1]{Eq.~(\ref{#1})}
\newcommand{\Figref}[1]{Fig.~\ref{#1}}

\begin{document}
\title{Analysis of STM images with pure and CO-functionalized tips: A first-principles and experimental study}
\date{\today}

\author{Alexander \surname{Gustafsson}}
\email{alexander.gustafsson@lnu.se}
\affiliation{Department of Physics and Electrical Engineering, Linnaeus University, 391 82 Kalmar, Sweden}
\author{Norio \surname{Okabayashi}}
\affiliation{Graduate School of Natural Science and Technology, Kanazawa University, 920-1192 Ishikawa, Japan}
\affiliation{Institute of Experimental and Applied Physics, University of Regensburg, D-93053 Regensburg, Germany}
\author{Angelo \surname{Peronio}}
\affiliation{Institute of Experimental and Applied Physics, University of Regensburg, D-93053 Regensburg, Germany}
\author{Franz J.  \surname{Giessibl}}
\affiliation{Institute of Experimental and Applied Physics, University of Regensburg, D-93053 Regensburg, Germany}
\author{Magnus \surname{Paulsson}}
%\email{magnus.paulsson@lnu.se}
\affiliation{Department of Physics and Electrical Engineering, Linnaeus University, 391 82 Kalmar, Sweden}

\begin{abstract}
We describe a first principles method to calculate scanning tunneling microscopy (STM) images, and compare the results to well-characterized experiments combining STM with atomic force microscopy (AFM). The theory is based on density functional theory (DFT) with a localized basis set, where the wave functions in the vacuum gap are computed by propagating the localized-basis wave functions into the gap using a real-space grid. Constant-height STM images are computed using Bardeen's approximation method, including averaging over the reciprocal space. We consider copper adatoms and single CO molecules adsorbed on Cu(111), scanned with a single-atom copper tip with and without CO functionalization. The calculated images agree with state-of-the-art experiments, where the atomic structure of the tip apex is determined by AFM. The comparison further allows for detailed interpretation of the STM images.
\end{abstract}

\maketitle

\section{Introduction}
Scanning tunneling microscopy (STM) and atomic force microscopy (AFM) have become standard tools to investigate surfaces and adsorbates 
on surfaces.  In addition to the atomic structure, STM can also characterize the electronic structure, shapes of molecular orbitals\cite{Repp2005,Gross2011}, vibrational\cite{Stipe1998}, and magnetic\cite{Hirjibehedin2007} excitations. Structures on the surface, such as adatoms and adsorbate molecules, can be manipulated by a STM tip\cite{Eigler1990}, and the force required to manipulate these structures 
can be determined by non-contact AFM\cite{Ternes2008,Emmrich2015prl}. The same technique can also be used to characterize the atomic structure of the very apex of the tip: when the tip is scanned above a CO molecule, one or more minima are observed in the frequency-shift image. Those minima correspond to the atoms composing the tip apex\cite{Welker2012,Emmrich2015}.

The structure of the tip apex strongly influences the tunneling processes. For example, we have recently demonstrated that the inelastic tunneling signal from a CO molecule on a Cu(111) surface is increased by using a sharp metallic tip whose apex consists of a single atom\cite{Okabayashi2016}. 
This increase is caused by the greater fraction of the tunneling electrons which passes trough the CO molecule. On the contrary, a blunter tip tunnels more electrons directly into the substrate, bypassing the molecule altogether.
Another example where the structure of the tip apex influences the tunneling process is the functionalization of the tip by a molecule. 
For a pentacene molecule adsorbed on an insulating layer, a metallic tip can be used to image a molecular orbital\cite{Repp2005}, whereas a CO-functionalized tip images a lateral derivative of the same\cite{Gross2011}.

To model STM measurements, the standard methods of Bardeen\cite{Bardeen1961} and Tersoff-Hamann\cite{Tersoff1985} have traditionally been used. The conceptually simple Tersoff-Hamann approach, that can be obtained from the Bardeen's approximation with an $s$-wave tip\cite{Hofer2003}, has provided a clear understanding of many STM experiments with non-functionalized STM tips\cite{Persson2002}. For CO functionalized STM, the Bardeen method\cite{Paz2006,Rossen2013,Bocquet1996,Teobaldi2007,Zhang2014}, the Chen's derivative rule\cite{Chen1990, Mandi2015, Mandi2015b}, and Landauer-based Green's function methods \cite{Cerda1997a,Cerda1997b} include the effects of multiple tip states. However, the understanding and {\it ab initio} modeling of STM experiments with and without chemical functionalization is still evolving. 

Recently, we developed an {\it ab initio} method based on the Bardeen's approximation for a CO molecule on a Cu(111) surface measured by a Cu tip with and without CO functionalization\cite{Gustafsson2016}. The wave functions close to the atoms were found in a localized basis set, whereafter they were propagated into the vacuum region in real space using the total DFT potential. The calculation in the $\Gamma$-point ($\bold{k}=\bold{0}$) 
reproduced qualitatively the dip in tunneling current for the CO/Cu(111). In this work, we extend the method to include $\bold{k}$-point sampling, and we calculate constant-height STM images of CO molecules and adatoms on a Cu(111) surface with a metallic and a CO-functionalized tip. 
The numerical calculations are compared to well-characterized STM experiments, where the structure of the tip apex is determined by AFM. An intuitive 
interpretation of the computational results is also presented in terms of the symmetry of the propagated wave functions between the tip and the substrate.

\section{Methodology}
\subsection{Theoretical}
To enable quantitative comparison between experiment and theory we have improved and extended our previous method\cite{Gustafsson2016} to simulate STM images from localized basis DFT calculations. We will provide a brief overview of the method, focusing on the main improvement: the inclusion of $\bold{k}$-point sampling. Details important for converged results are also presented.

\begin{figure}[tb]
	\includegraphics[width=\columnwidth]{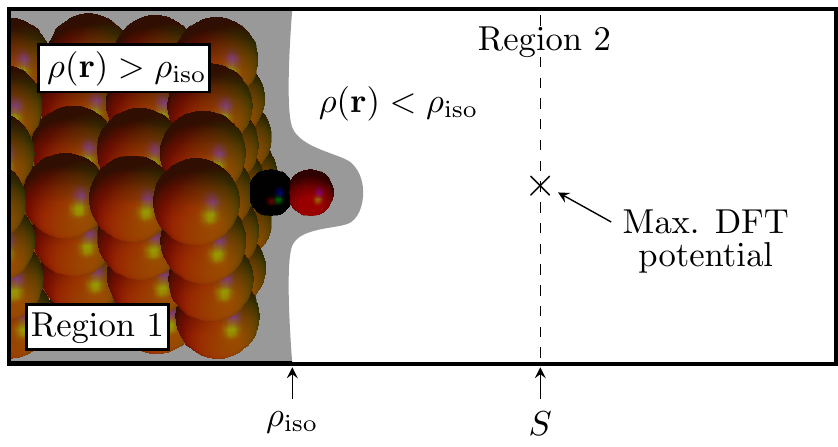}
	\caption{
	Propagation of the wave functions into region 2 is performed by matching the localized-basis wave functions, $\psi_{t,s}^k$, from region 1 on the electron charge density isosurface, $\rho_{\textrm{iso}}$, and solving for region 2 using a finite difference real space grid. To the right of the separation surface, $S$, the potential is the average of the potential at $S$, in order to simulate a constant vacuum potential far from the surface.}
	\label{figCartoon}	
\end{figure}

In the Bardeen's approximation\cite{Bardeen1961}, the transmission coefficient for a given $\bold{k}$-point at the Fermi energy reads $T^k(\varepsilon_{\textrm{F}})=\sum_{ts} \, 4\pi^2|M^k_{ts}|^2$, where the sum is limited to the states at the Fermi energy, i.e., the integration over energy has already been carried out to remove the delta functions sometimes included in the formalism\cite{Hofer2003}. The matrix elements $M^k_{ts}$ are
\begin{widetext}
\beq
M^k_{ts}=-\frac{\hbar^2}{2m}\int_S\textrm{d}S\,\left[{\varphi^k_t}^*(\bold{r}+\bold{R})\,\nabla\varphi^k_s(\bold{r})-\varphi^k_s(\bold{r})\,\nabla{\varphi^k_t}^*(\bold{r}+\bold{R})\right],\label{eqMts}
\eeq 
\end{widetext}
where $\varphi^k_{t,s}$ are the wave functions using the indices $t,s$ to 
sum the states from the tip/substrate. The integral (over $\bold{r}$) is performed 
on a surface intersecting the vacuum region, which in our case is chosen as a flat surface at the position where the total potential (from DFT) has its maximum, see \Figref{figCartoon}. 

Localized basis  DFT calculations cannot accurately describe the 
wave function in the vacuum region since the basis set is localized on the atoms. We therefore use the DFT wave functions (calculated according to the theory outlined in \cite{Paulsson2007}), $\psi^k_{t,s}$, as boundary conditions on a surface close to the tip/substrate, where these wave functions are described accurately by localized basis, see \Figref{figCartoon}. The surface from which $\psi_{t,s}^k$ are propagated, is an iso surface of constant charge density, computed by DFT, as the sum of the absolute value squared of all Kohn-Sham orbitals up to the Fermi energy. We then solve for the wave functions  in the vacuum region using a finite difference (FD) grid, taking into account the total potential obtained from the DFT calculation. For a given energy and $\bold{k}$-point, this corresponds to solving a sparse linear system of equations, which can be performed efficiently. These real-space wave functions, $\varphi^k_{t,s}$, are computed separately from the substrate ($s$) and tip ($t$), whereafter they are combined, \Eqref{eqMts}, at the separation surface $S$. Furthermore, the STM image can be obtained by scanning $\bold{R}$ in \Eqref{eqMts}, i.e., translating the tip wave functions laterally over the substrate. This convolution is computationally cheap when performed with fast Fourier transforms\cite{Paz2006,Zhang2014}. Using this method we can investigate systems containing several hundred atoms within a reasonable timeframe. That is, the systems studied here were performed in a few days for the complete geometry optimization, wave functions calculation, propagation, and STM simulation using $\le16$ cores.

Both the DFT calculation and FD propagation use periodic boundary conditions along the surface ($x$-$y$) plane. This supercell approach is in keeping with computational traditions and provides correct results for large supercells. Our previous investigations\cite{Gustafsson2016} were restricted to the $\Gamma$-point ($\bold{k}=\bold{0}$) which we here extend to include $\bold{k}$-point sampling in the $x$-$y$ plane. The wave function calculations are performed for each $\bold{k}$-point, and the computational time therefore scales linearly with the number of $\bold{k}$-points. The final STM image is then computed as the average over $\bold{k}$-points. As shown below, $\bold{k}$-point sampling is crucial to obtain quantitative results.

Neither bias-voltage dependence nor constant-current STM images are at present provided in the theory. The former may be achievable via a self-consistent calculation of a non-equilibrium potential, whereafter the scattering states close to the surface at each side of the vacuum region are calculated at respective energy level. However, bandgap underestimation by DFT will certainly affect bias-voltage STM spectroscopy in addition to modifying the low bias conductance. This might be alleviated by the use of more advanced DFT/hybrid functionals. A constant-current mode is also straightforward to implement, and possible to compute from a single STM calculation, provided that the considered vacuum region is large, and the tip-height variation (along $\hat{z}$) is not too large. Since the propagated wave functions are computed separately from each side, it is for instance possible to store the tip-wave functions, and scanning a specific surface adsorbate species immediately with different tips, provided that the lateral unit cell has the same dimensions. Hence, using different materials in the substrate and tip may currently be difficult, since periodicity in the lateral plane is crucial. 

\subsection{Computational details}
We have used the \textsc{Siesta}\cite{Soler2002}  DFT code to compute geometry optimization on a slab consisting of eight Cu layers where each layer has $6\times6$ Cu atoms with a nearest-neighbour distance of 0.257 nm. The lateral cell dimensions are therefore $1.54\times1.54$ nm. This size corresponds to a maximal tip-substrate distance that can be used in the calculation to avoid large influences of the next lateral unit cell. However, there also exists a minimal tip-substrate distance due to the approximation that the states of the substrate are unaffected by the potential from the tip and {\it vice versa}. The tip-substrate distance is defined as the vertical distance between the outermost tip apex atom and the top Cu surface layer of the substrate. As shown previously\cite{Gustafsson2016}, the minimum distance for the systems considered here is approximately 0.8 nm. In the calculations presented below we use a tip-substrate distance of $1.21(1.01)$ nm for the pure Cu tip (CO functionalized tip).

The  \textsc{Siesta} calculations were performed using the Perdew-Burke-Ernzerhof (PBE) parametrization of the generalized gradient approximation (GGA) exchange-correlation functional\cite{Perdew1996}, double-(single-) zeta polarized basis set for C, O (Cu) atoms, a 200 Ry real space mesh cutoff and $4\times4$ $\bold{k}$-points. The Cu atoms at the substrate surface, and the tip, have longer radial range (by 2~\AA) compared to Cu atoms in the bulk/electrodes, which have $4.3$~\AA~range. This assures that adsorbates that protrude much from the surface, e.g., CO/adatom/Cu(111), give accurate STM images, due to the contribution from the Cu surface beneath the adsorbate. The STM tip is represented by a pyramidal tip consisting of four Cu atoms on one side of the slab while the molecule is placed on the opposing side. The geometry optimization concerns the adsorbate molecule, the two top substrate layers, and the tip atoms (forces less than 0.04 eV/\AA). Nine additional Cu layers are thereafter added, and the wave functions are calculated by \textsc{Transiesta}\cite{Brandbyge2002} and \textsc{Inelastica}\cite{frederiksen2007}. Note that in the calculations of the substrate/tip wave functions we consider approximately 600 atoms in the DFT calculations.

The PBE-GGA functional provides an energy minimum for the CO molecule at a hollow site at the Cu(111) surface\cite{Feibelman2001}, whereas experiments show that a top site is the most stable one. In this work the CO is simply initially situated in vicinity of a top site, whereafter a geometry optimization is performed, so that a local energy minimum is found at a top site. Different adsorption sites, and their influence on STM images, are not further investigated. 

For the real-space wave function propagation we have found that the real space grid given by \textsc{Siesta} (200 Ry cutoff) is  unnecessarily fine, and in the wave function propagation the grid coarseness is doubled (corresponding to a 50 Ry cutoff in \textsc{Siesta}) in the lateral plane while unchanged in the transport direction. This reflects that the potential and the wave functions in the vacuum region change slowly compared to closer to the nuclei. The down-sampled grid size consist of $60\times60$ points (lattice constant $\sim0.25$ \AA) in the plane of the substrate and approximately 150 points along the transport direction, meaning that the discrete Laplacian matrix has dimensions of approximately $0.5\textrm{M}\times0.5\textrm{M}$. The electron density at the isosurface, $\rho_{\textrm{iso}}$, which divides the space into two regions, is chosen so that the surface lies well outside the radii of the pseudopotentials, and well inside the range of the 
localized basis orbitals. Incidentally, the isosurface lies close to where the Fermi energy crosses the total potential. Henceforth, 
the isovalue $\rho_{\textrm{iso}}=10^{-3}$ [Bohr$^{-3}$Ry$^{-1}$] is used in all calculations. 

\begin{figure}[tb]
	\includegraphics[width=\columnwidth]{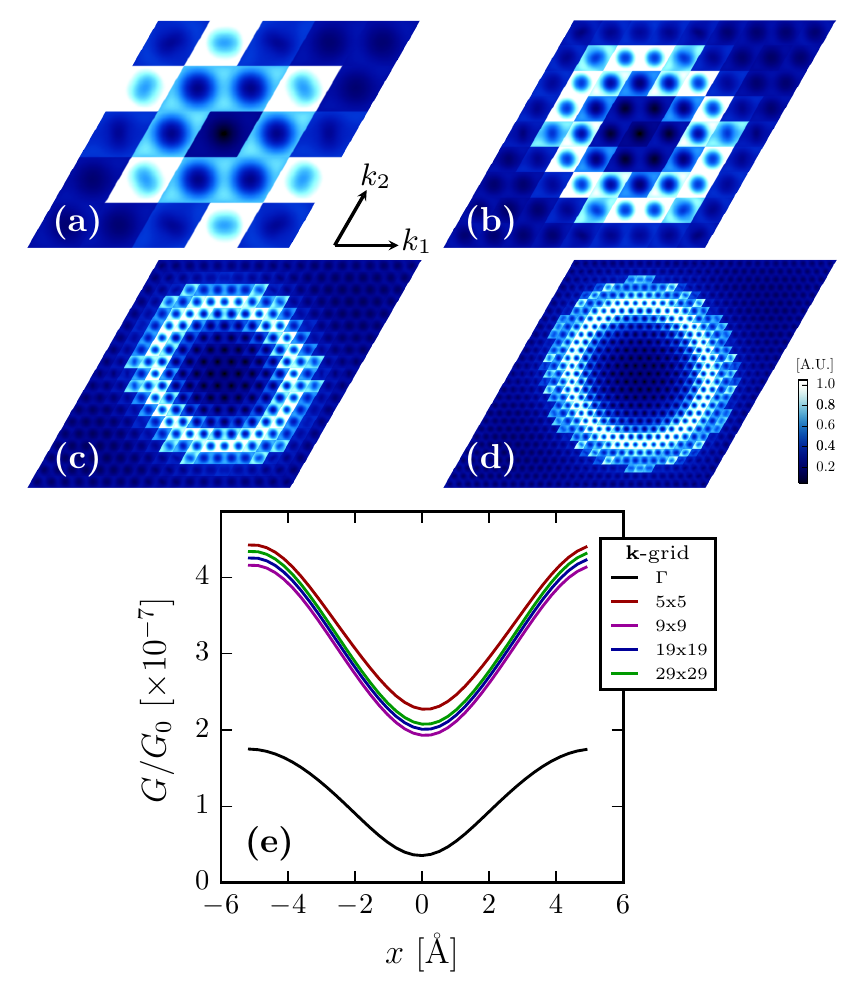}
	\caption{Calculations showing STM images for all individual $\bold{k}$-points using a (a) $5\times5$, (b) $9\times9$, (c) $19\times19$, and (d) $29\times29$ $\bold{k}$-grid for the CO/Cu(111) system with the smaller $4\times4$ surface supercell. 
	The tip apex is scanned 1.06 nm above the copper substrate, i.e. 0.75 nm above the oxygen atom of the adsorbed CO.
	Figure (e) shows the $\bold{k}$-averaged cross section STM images in (a)-(d), and also includes the $\Gamma$-point image. The $\bold{k}$-points are homogeneously distributed in the reciprocal space of the supercell.} \label{figBZ}	
\end{figure}

\subsection{$\bold{k}$-point convergence}
We have found that a converged $\bold{k}$-point sampling is essential for detailed comparison with experiments. To illustrate the STM images for different $\bold{k}$-points we show the calculated STM images for the CO molecule with a Cu tip for each $\bold{k}$-point in \Figref{figBZ}. To speed up the calculations here, we use layers of $4\times4$ Cu atoms, since we have confirmed that also a $4\times4$ surface yields a quantitative agreement to experiment with a sufficient number of $\bold{k}$-points\footnote{Other systems (adatom, and CO/adatom scanned by a Cu tip, and adatom scanned by a CO tip) have proved to demand the larger $6\times6$ surface for quantitative comparison to experiment.}. That is, the size of the lateral unit cell and tip-substrate distance are decreased compared to the results below, while keeping the remaining parameters unchanged. As shown in the figure, there is a significant difference between the $\Gamma$-point and the $\bold{k}$-point averaged results, both in the magnitude of the tunneling current and the min/max ratio. However, for this system the $\Gamma$-point image yields a qualitative similarity to the experiment, while for other systems even the qualitative shape is changed compared to a converged $\bold{k}$-averaged image. An example of this behaviour is the CO at the Cu(111) surface scanned by a CO terminated tip, where the $\Gamma$-point calculation gives a peak\cite{Gustafsson2016}, whereas its $\bold{k}$-converged image gives a dip, see \Figref{FigTheoryCO} (a) and (c). Although for this system the cross sections seem to have converged already for $5\times5$ $\bold{k}$-points, other combinations of substrate/tip shown below demand an even higher number of $\bold{k}$-points. In the calculations using the $6\times6$ layers presented above and below, we have used $11\times11$ $\bold{k}$-points to ensure convergence.

\subsection{Experimental details}
The experiments are performed with an ultra-high vacuum low-temperature (4.4 K) STM and AFM combined machine (LT-STM/AFM, Scienta Omicron, Taunusstein, Germany) located at Regensburg University. The (111) surface of a copper single crystal is cleaned by repeated sputtering and annealing cycles, before being loaded into the microscope. CO molecules are adsorbed \emph{in situ} by backfilling the vacuum chamber. An etched tungsten wire is adopted as a tip attached to a force sensor\cite{Giessibl2000}, which is repeatedly poked into the Cu substrate to prepare a sharp tip whose apex consists of a single atom, as confirmed by the frequency shift image of the tip scanned above a CO molecule\cite{Welker2012,Emmrich2015}. The tip apex is probably coated by Cu atoms owing to the repeated poking processes (see supplemental material in Ref.~\cite{Hofmann2014}). This also scatter Cu adatoms on the Cu(111) substrate\cite{Emmrich2015} which we image before and after having adsorbed a CO molecule on them by atomic manipulation\cite{Okabayashi2016}. All the STM images presented here are acquired at constant height, oscillating the sensor with an amplitude $A=20$ pm. The measured tunneling current $I_t$ is thus the average over the sensor oscillation\cite{Majzik2012,Huber2015}.

\section{Experimental results}

\begin{figure}[tb]
	\includegraphics[width=\columnwidth]{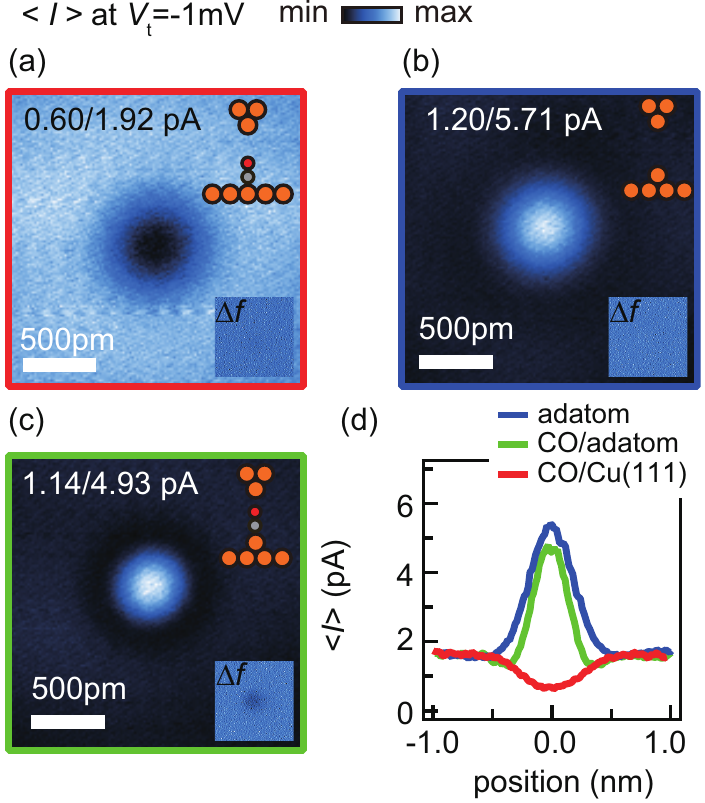}	
	\caption{Experimentally observed constant-height current images of (a) a CO molecule, (b) a Cu adatom, (c) a CO molecule on a Cu adatom, all of which are adsorbed on the Cu(111) surface and are measured by a single atom tip. The set-point is $V_t=-1$ mV and $I_t=1.7$ pA on the Cu(111) surface for all cases. The insets show the frequency shift image simultaneously measured with the current image. (d) The cross-sections of the constant-height current images shown in (a)-(c). } \label{FigExpCu}
\end{figure}

Figure \ref{FigExpCu} shows constant-height current images for (a) a CO molecule, (b) a Cu adatom and (c) a CO molecule on a Cu adatom, all adsorbed on a Cu(111) surface, and scanned by a single atom tip. For the three cases, an identical set-point on the Cu(111) surface is adopted (sample bias $V_t=-1$ mV and $I_t=1.7$ pA), where the interaction between the tip and sample is confirmed to be a negligible attractive force (see the insets in \Figref{FigExpCu} (a)-(c)). The cross-sections of the three images are shown in \Figref{FigExpCu} (d) where the current approaches the set-point far from the molecule. We see that the current for the CO molecule on the Cu (111) surfaces is 37\% of that on the Cu(111) surface\footnote{The ratio of the current on the CO molecule to the Cu(111) surface is decreased to 23\% when the tip locates closer to the CO molecule by 335 pm ($V_t=-1$ mV and $I_t=1.5$ nA on Cu(111) surface), which is consistent with the previous report by single atom tips with enough sharpness\cite{Okabayashi2016}.}. On the other hand, the current on the Cu adatom is 3.2 times higher than that on the Cu(111) surface. Comparing the current on the CO molecule and that on the Cu adatom, the former is 12\% of the latter: a CO molecule is one order of magnitude less conductive than a Cu adatom when adsorbed on a Cu(111) surface. When a CO molecule is adsorbed on a Cu adatom, the image on the CO molecule shows a bright spot. However, the current on the molecule is decreased to 89\% of that on the Cu adatom, i.e., the current on a CO molecule is again decreased compared to the case without the CO molecule, similarly to the case of the Cu(111) substrate. Decreasing the tip-sample distance by 200 pm does not change these features.

\begin{figure}[tb]
	\includegraphics[width=\columnwidth]{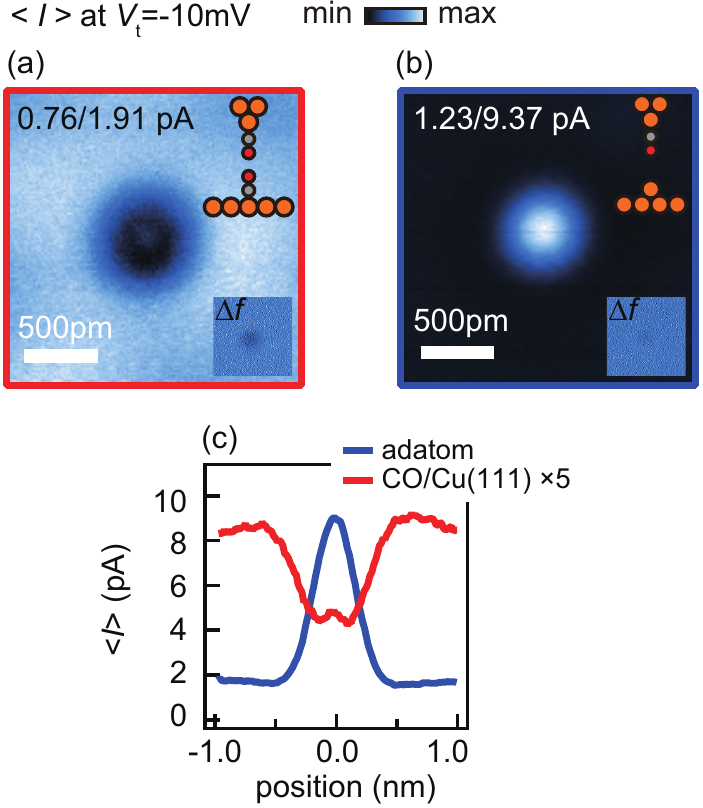}	
	\caption{Experimentally observed constant-height current images of (a) a CO molecule and (b) a Cu adatom adsorbed on the Cu(111) surface measured by a CO functionalized tip. The set-point is $V_t=-10$ mV and $I_t=1.7$ pA on the Cu(111) surface for both cases. (c) The cross-sections of the constant-height current images in (a) and (b). } \label{FigExpCO}
\end{figure}

The lower conductivity of a CO molecule is also observed with a CO functionalized tip\cite{Bartels1997}. When the CO functionalized tip is scanned above a CO molecule and a Cu adatom on the Cu(111) at a small interaction set-point ($V_t=-10$ mV and $I_t=1.7$ pA on the Cu(111) surface), the current on the CO molecule is decreased to 11\% of that on the Cu adatom (see \Figref{FigExpCO} (a)-(c)). In addition, we see the striking difference in the image of the CO molecule comparing to the case by the single atom tip in \Figref{FigExpCO} (a): the current image just above the CO molecule shows a small peak, which is centered in the surrounding dip similarly observed in the single atom tip image. This difference should originate from the more prominent $p$ states of the tip, owing to the CO functionalization\cite{Gross2011}, which will be discussed later in the comparison with the theory. Owing to this effect, the current on the CO molecule is 58\% of that on the Cu(111) surface at this set-point.

\begin{figure}[tb]
	\includegraphics[width=\columnwidth]{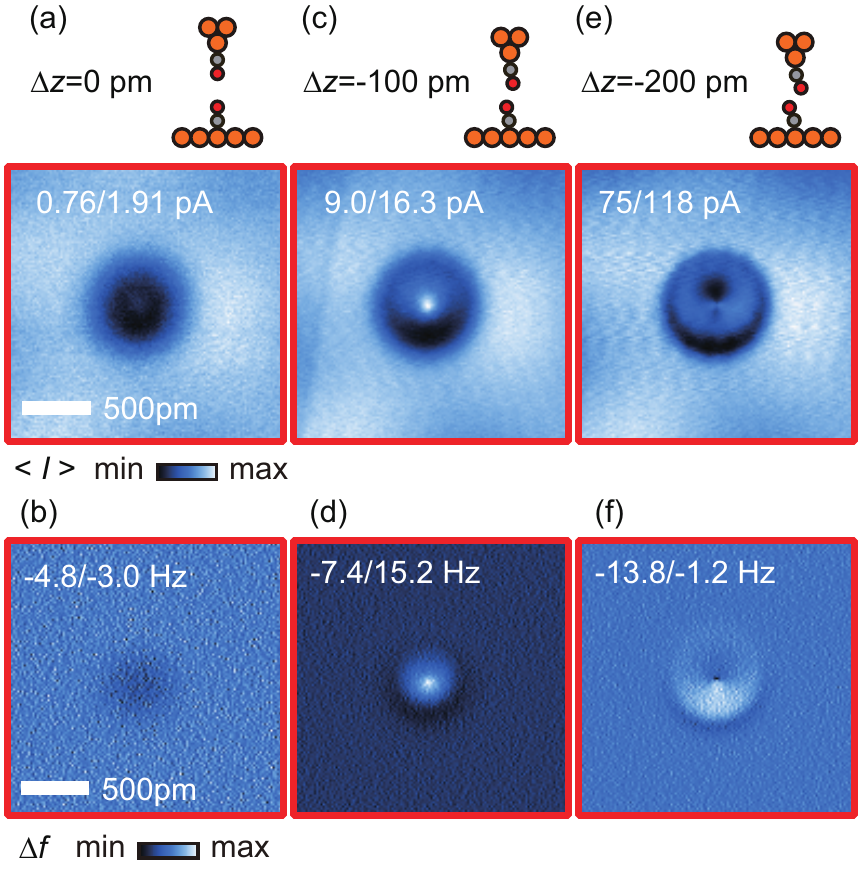}	
	\caption{Experimentally observed constant-height current and frequency shift images of a CO molecule on a Cu(111) surface by a CO functionalized tip for various tip sample distances at $V_t=-10$ mV. The tip position is distant in (a)(b), middle in (c)(d), and near in (e)(f). } \label{FigExpFreq}
\end{figure}

The feature of the STM image with the CO tip for the Cu adatom is constant when the tip-sample distance is decreased by 200 pm, while this is not the case for the CO molecule on the substrate, see \Figref{FigExpFreq}. When the tip position is far from the surface CO molecule ($\Delta z=0$ pm), the frequency shift image shows a small attractive feature between the two CO molecules (see \Figref{FigExpFreq} (b)), which should keep the alignment of two CO molecules parallel\cite{Weymouth2014,Sun2011}. This parallel alignment can keep the relative symmetry between the $p$ states of the tip and that of the substrate and thus increase the tunneling current via the two $p$ states. On the other hand, when the tip position is 100 pm closer to the CO molecule on the surface, the repulsive feature appears in the frequency shift image (see \Figref{FigExpFreq} (d)), which results in a small CO bending by the repulsive force between the two CO molecules\cite{Weymouth2014,Sun2011}. This small CO bending should enhance the direct tunneling into the Cu substrate and is probably the origin of the enhanced bright spot in the center of the STM image in \Figref{FigExpFreq} (c). Further reduction of the tip sample distance ($\Delta z=200$ pm), means a strong repulsive interaction (see \Figref{FigExpFreq} (f)), which results in a large CO bending, and by which the current image is strongly deformed as shown in \Figref{FigExpFreq} (e). The present result indicates that the $p$ state interaction and the CO bending are the two candidates for the origin of the central bright spot, which can be discriminated by combining AFM and STM.

\section{Theoretical results}
\subsection{Computational results}
\begin{figure}[tb]
	\includegraphics[width=\columnwidth]{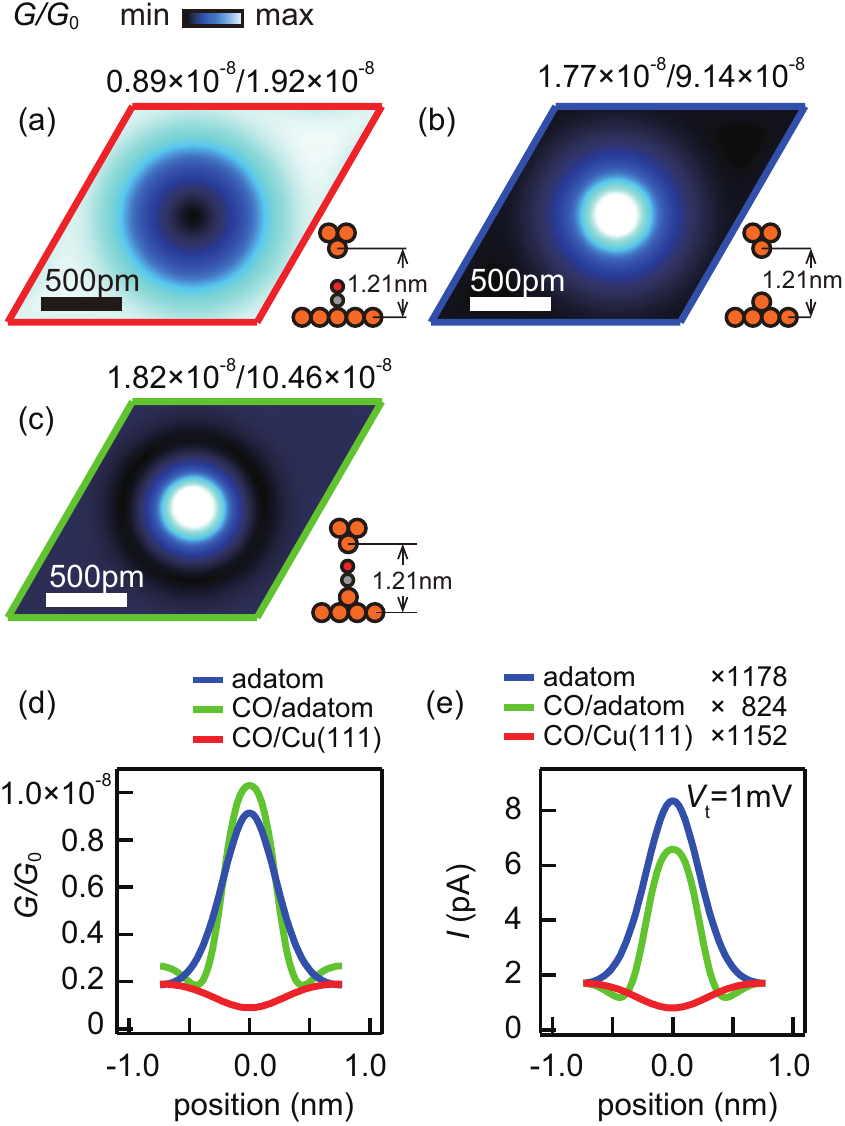}	
	\caption{Theoretically calculated constant-height conductance images of (a) a CO molecule, (b) a Cu adatom, (c) a CO molecule on a Cu adatom, all adsorbed on a Cu(111) surface. (d) The cross-sections of the conductance images shown in (a)-(c). (e) The calculated conductances are scaled such that the current on the Cu(111) is identical to the experimental value. } \label{FigTheoryCu}
\end{figure}

Calculations for the system adopted in \Figref{FigExpCu} are shown in \Figref{FigTheoryCu}, where the distance between the tip apex and the top Cu substrate layer is 1.21 nm for all cases. This large distance, more than 0.7 nm from outermost adsorbate atom to tip apex, is chosen to ensure that the potential of the tip does not affect the wave functions from the substrate and {\it vice versa}. The computational calculation provides the conductance at the Fermi energy (see  \Figref{FigTheoryCu} (a)-(d)), which has been converted to the current $I_t$ at $V_t=1$ mV by using the formula $I_t=G V_t$ so that comparison to experiments is possible. To compare absolute current levels, a scaling factor is employed to set the Cu substrate current consistent with the experiment: $1.7$ pA at $\left| V_t \right|=1$ mV (see  \Figref{FigTheoryCu} (e)).

\begin{figure}[tb]
	\includegraphics[width=\columnwidth]{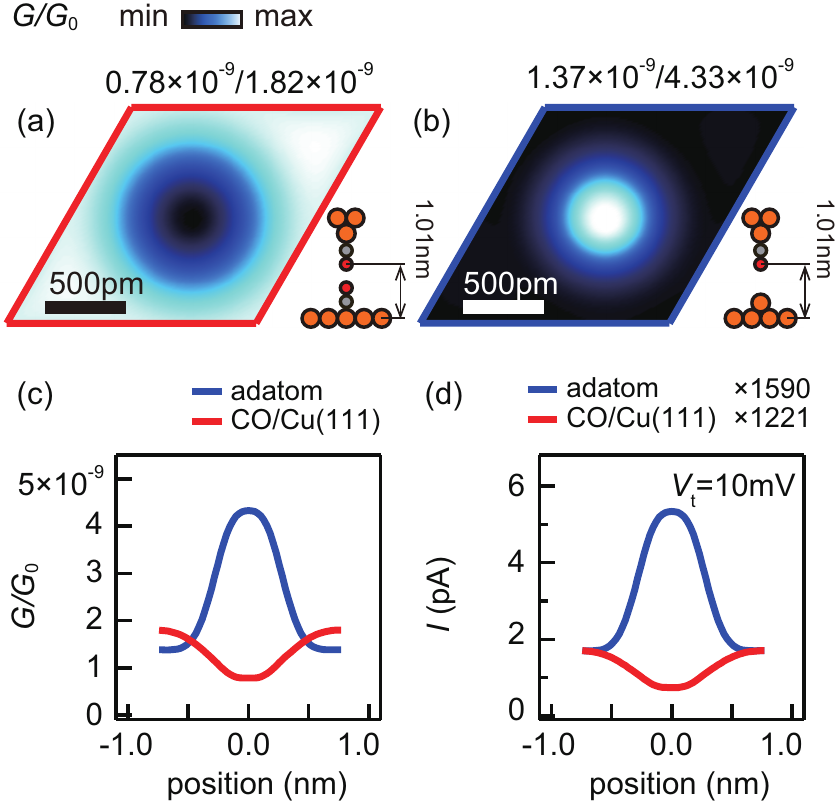}	
	\caption{Theoretically calculated constant-height conductance images of (a) a CO molecule and (b) a Cu adatom adsorbed on the Cu(111) surface for a CO functionalized tip. (c) The cross-sections of the conductance images shown in (a)(b). (d) The calculated conductance is scaled such that the current on the Cu(111) is identical to the experimental value.}  \label{FigTheoryCO}
\end{figure}

The main results of the theory can be summarized as follows. (1) the current image of a CO on a Cu(111) surface shows a dark spot where the current on the CO molecule is 46\% of that on the Cu(111) surface. (2) The Cu adatom shows a bright spot whose current is 5.2 times higher than the value on the Cu(111) substrate. Comparing the current on the CO molecule and that on the Cu adatom, the former is 9\% of the latter. (3) The image of a CO molecule adsorbed on a Cu adatom shows a bright spot, where the current, after the multiplication of the scaling factor, is 79\% of that on the Cu adatom. This summary is consistent with the experimental findings in  \Figref{FigExpCu}: (a) a CO molecule is less conductive than a Cu adatom on a Cu substrate by one order of magnitude, and (b) the current on a CO molecule is decreased comparing to the case without the CO molecule. In addition, the theory reproduces the small current depression around the central peak for the CO/adatom system, see \Figref{FigTheoryCu} (e).

The same consistency between theory and experiment can be seen for the case of the CO functionalized tip (see  \Figref{FigTheoryCO}). In accordance with the case of  \Figref{FigExpCO}, a CO molecule and a Cu adatom adsorbed on the Cu(111) surface is simulated with a CO functionalized tip, where the distance between the O atom in the tip and the top Cu substrate layer is 1.01 nm for both cases. The current on the CO molecule after the multiplication of the scaling factor is 14\% of that on the Cu adatom: the smaller conductivity of a CO molecule compared to a Cu adatom is again reproduced for the CO functionalized tip. 

\begin{figure}[tb]
	\includegraphics[width=\columnwidth]{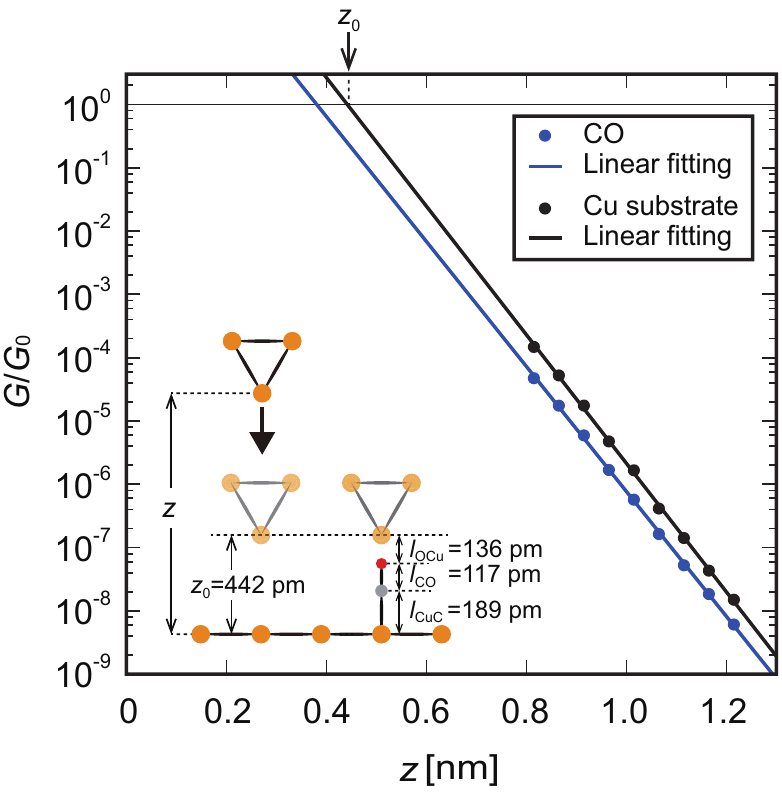}
	\caption{Calculated conductances for a series of tip-substrate distances showing the exponential decay of the conductance. Since the slopes are slightly 
	different close to the substrate for the conductance over the molecule/substrate, the relative depth of the dark spot will depend weakly on the tip-substrate 
	distance.} \label{figExpDecay}	
\end{figure}

We further investigate the influence of the tip-substrate distance by simulating the CO-substrate/Cu-tip system with several different tip-heights, see \Figref{figExpDecay}. Overall the exponential decay ($e^{- 2 \kappa z}$) shows the same behaviour over the substrate and molecule with only a small difference closer to the substrate 
with exponential decay factor $\kappa = 11.4(11.7)$/nm over the molecule(surface). These values are consistent with our experimental results of $\kappa\approx10$/nm for both over molecule and substrate. The exponential decay should approach the work function ($W$) far from the substrate, $\kappa =  \sqrt{2 m W}/\hbar$. The fitted theoretical $\kappa$ corresponds to a work function of $5.0\pm0.2$ eV which agrees with previous calculations (5.3 eV)\cite{Skriver1992} and experimental results (5.0 eV)\cite{Michaelson1977}. We also note that $e^{2 \kappa\times\textrm{0.1 \mbox{nm}}} \approx 10$, i.e., the conductance changes by approximately an order of magnitude per 1 \AA.

If the exponential decay over the Cu substrate is extrapolated to smaller tip-substrate distances, the conductance reaches the conductance quantum at $z_0=442$ pm over the substrate, corresponding to 136 pm above the O atom. The position $z_0$ has frequently been adopted as the origin of the vertical position for the force measurements\cite{Ternes2008,Emmrich2015prl,Welker2012}. We note that in the extrapolation to shorter tip-substrate distances, we do not account for any
displacement of the tip-apex atom and substrate atoms beneath the tip. This can cause the experimental $I$-$V$ curve to deviate from the exponential dependence\cite{Olesen1996}.

\subsection{Interpretation}

\begin{figure}[tb]
	\includegraphics[width=\columnwidth]{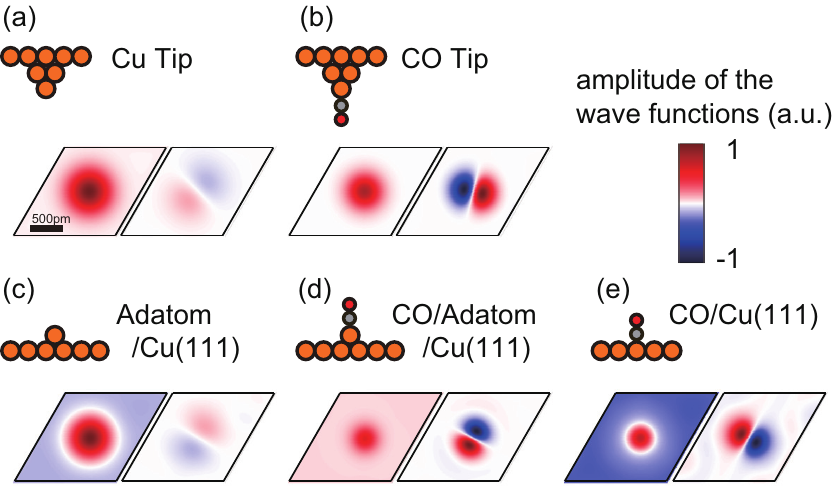}	
	\caption{Calculated $\Gamma$-point wave functions in the vacuum gap from (a) a Cu tip, (b) a CO tip, (c) a Cu adatom, (d) a CO on a Cu adatom and (e) a CO on a Cu(111) surface. To qualitatively explain the relationship between the wave functions and the tunneling current, the $s$- and one of the two degenerate $p$ states are selected.\label{FigIntuitive}} 
\end{figure}

The results of the numerical calculations can be intuitively interpreted by considering the schematic images of the wave functions, see  \Figref{FigIntuitive}, over (a) a Cu tip, (b) a CO tip, (c) a Cu adatom, (d) a CO on a Cu adatom and (e) a CO on the Cu(111) 
surface. Since the calculations involve approximately 30 wave functions for each $\bold{k}$-point\footnote{This means $\mathcal{O}(10^3)$ tip-substrate combinations for each $\bold{k}$-point, which overall gives $\mathcal{O}(10^5)$ combinations that contribute to the total current.}, we only show the most prominent states from the $\Gamma$-point calculations in order to simplify the presentation\footnote{A full set of wave functions at the $\Gamma$-point can be found in Ref.~\cite{Gustafsson2016} for some of the systems.}. These tip- and substrate wave functions have large amplitudes, and the most conductive tip-substate combination typically gives 10-20\% of the total current. The plane on which the wave functions are drawn is the integration surface close to the middle of the vacuum gap. For the purpose of an intuitive interpretation, we have labeled the wave functions according to their symmetry in the lateral plane as $s$- or $p$ states, where only one of the two degenerate $p$ states are shown, i.e., the 90 degree rotated $p$ state is omitted. Note that the direction of the symmetry axis for the $p$ state depend on minute numerical details and is therefore not the same for the different systems. 

The features of the wave functions in \Figref{FigIntuitive} can be summarized as follows. In the case of the Cu tip, the main weight of the wave functions is of $s$-type which is centered on the apex atom with a negligible amplitude over the Cu substrate, 
see \Figref{FigIntuitive} (a). The wave functions for the Cu adatom are similar to those for the Cu tip, however, we see a slight difference in the $s$-type wave function: the absolute value of the amplitude on the Cu substrate is slightly increased up to 14\% of that on the Cu adatom with a sign change between over the Cu adatom and over the Cu substrate (\Figref{FigIntuitive} (c)). This difference comes from the fact that the Cu adatom is one layer closer to its Cu substrate compared to the Cu-tip apex atom (due to the four-atom pyramid tip). Hence, some contribution from the Cu surface is visible in \Figref{FigIntuitive} (c). When a CO molecule is adsorbed on a Cu adatom or the tip apex, a similar tendency is observed for the $s$-type wave functions (see \Figref{FigIntuitive} (b) and (d)): the amplitude on the Cu substrate is negligible for the CO tip while it has a small amplitude for the CO molecule on the Cu adatom. In addition, the relative amplitude of the $p$ states increase in both cases. In contrast, for the CO molecule adsorbed on the clean Cu substrate, the amplitude of the $s$ state over the substrate is substantial (49\%) compared to the value over the CO molecule, see \Figref{FigIntuitive} (e). The sign change of the the $s$ wave function when moving from the substrate to over the molecule is important for the further discussion.

The matrix element of the tunneling current $M_{ts}$ can qualitatively be approximated by the surface integral of the product of the two wave functions from the tip and the substrate, see Eq.~\ref{eqMts} and Ref.~\cite{Gustafsson2016}. Judging from the symmetry of the two wave functions, an $s$-$p$ combination gives a negligible contribution when the tip apex is aligned with the adsorbate.  We will therefore discuss the $s$-$s$ and the $p$-$p$ combinations. The latter seems relevant only when a CO molecule is present both on the tip and on the surface.

For the Cu tip, the states are mainly of $s$-type with a negligible amplitude away from the tip. When scanning this tip over the Cu adatom, whose wave 
function has a small amplitude on the Cu substrate, the current shows  a maximum with the tip positioned over the adatom (with a small current depression surrounding the central peak due to the slight sign change of the substrate $s$ wave). The same is true for the 
CO molecule on the adatom, i.e., the $s$ waves of the tip/substrate both have a small amplitude on the substrate. Thus, the image of the CO molecule on the Cu adatom becomes bright. On the other hand, when the tip is scanned over the CO molecule on the Cu(111) surface, the substrate $s$ state is substantial over the bare Cu substrate, which provides a large current even when the tip is not centered on the molecule. 
In addition, the change of sign when moving from the substrate to above the adsorbed CO decreases the surface integral by cancellation when the tip is centered on the molecule. The CO molecule therefore appears as a dark spot on Cu(111) as shown in \Figref{FigTheoryCu} (a). A fully equivalent statement is that the decrease in the current over the molecule is caused by the interference between tunneling through the vacuum gap and through the molecule, i.e., the current pathways through space and through the molecule have different signs\cite{Nieminen2004}.

Next we consider the CO functionalized tip which has similar dominating states as the CO on the adatom with a single-sign $s$ state and prominent $p$ states. When the tip is scanned over the Cu adatom, the surface integral becomes large when the tip is right above the adatom, resulting in a bright spot in the current image. On the other hand, when the tip is scanned over the CO molecule adsorbed on the Cu(111) surface for which the $s$ state changes sign over the molecule, the $s$-$s$ surface integral becomes small and, similarly to the Cu tip, results in a lower current over the molecule. However, in this case both sides have moderate amplitudes in the $p$ states. The $p$-$p$ contribution is therefore significant and is responsible for the observed bright spot in the $\Gamma$-point\cite{Gustafsson2016}. This may be another reason (besides the possibly tilted bonding angle) to the experimentally observed centered structure seen in the surrounding dip in \Figref{FigExpCO} (c). We stress that the previous discussion, regarding the most prominent $\Gamma$-point wave functions, only provides a qualitative sketch of the STM images.

\section{Summary}
We have presented a first principle method for calculating STM images based on localized basis DFT. The method is computationally efficient and can be applied to large systems. In addition, tip states are treated on the same footing as substrate states and can therefore model chemically functionalized STM tips.  We have further provided experimental benchmark measurements for a series of CO structures on Cu(111) with both normal and CO functionalized STM tips. The STM measurements are well characterized including examination of the tip apex structure by frequency modulated AFM. The comparison between the experiment and theory shows near quantitative agreement and the qualitative features were elucidated from the theoretical wave functions.  

\section{Acknowledgements}
The computations were performed on resources provided by the Swedish National Infrastructure for Computing (SNIC) at Lunarc. A.G. and M.P. are supported by a grant from the Swedish Research Council (621-2010-3762). The experimental works were supported by funding (SFB 689) from Deutsche Forschungsgemeinschaft (F.J.G) and by JSPS KAKENHI Grant Number JP16K04959 (N.O.). Experimental support from Daniel Meuer and Alexander Liebig is gratefully acknowledged.

\bibliographystyle{apsrev}
\bibliography{/Users/aguadmin/Jobb/PhD/RefPapers/agref.bib}
\end{document}